# A two-component flavin-dependent monooxygenase involved in actinorhodin biosynthesis in *Streptomyces coelicolor*

Julien Valton, Laurent Filisetti, Marc Fontecave*, Vincent Nivière*

Laboratoire de Chimie et Biochimie des Centres Redox Biologiques, DRDC-CEA/CNRS/Université Joseph Fourier, 17 Avenue des Martyrs, 38054 Grenoble Cedex 9, France.

* To whom correspondence should be addressed. Vincent Nivière; Tel.: 33-4-38-78-91-09; Fax: 33-4-38-78-91-24; E-mail: vniviere@cea.fr. Marc Fontecave; Tel.: 33-4-38-78-91-03; Fax: 33-4-38-78-91-24; E-mail: mfontecave@cea.fr.

Running title : ActVB/ActVA, a two component FMN-dependent monooxygenase

## SUMMARY

The two-component flavin-dependent monooxygenases belong to an emerging class of enzymes involved in oxidation reactions in a number of metabolic and biosynthetic pathways in microorganisms. One component is a NAD(P)H:flavin oxidoreductase which provides a reduced flavin to the second component, the proper monooxygenase. There, the reduced flavin activates molecular oxygen for substrate oxidation. Here, we study the flavin reductase ActVB and ActVA-ORF5 gene product, both reported to be involved in the last step of biosynthesis of the natural antibiotic actinorhodin in *Streptomyces coelicolor*. For the first time, we show that ActVA-ORF5 is a FMN-dependent monooxygenase which together with the help of the flavin reductase ActVB catalyze the oxidation reaction. The mechanism of the transfer of reduced FMN between ActVB and ActVA-ORF5 has been investigated. Dissociation constant values for oxidized and reduced flavin ($FMN_{ox}$ and $FMN_{red}$) with regard to ActVB and ActVA-ORF5 have been determined. The data clearly demonstrate a thermodynamic transfer of $FMN_{red}$ from ActVB to ActVA-ORF5, without involving a particular interaction between the two protein components. In full agreement with these data, we propose a reaction mechanism in which $FMN_{ox}$ binds to ActVB where it is reduced and the resulting $FMN_{red}$ moves to ActVA-ORF5 where it reacts with $O_2$ to generate a flavin-peroxide intermediate. A direct spectroscopic evidence for the formation of such species within ActVA-ORF5 is reported.

## INTRODUCTION

There is a great variety of monooxygenases, catalyzing oxygen activation and oxygen transfer reactions in the living world. The most extensively studied ones are the cytochrome P450-dependent monooxygenases (1) and the enzymes belonging to the methane monooxygenase family (2), both using an iron center for $O_2$ activation. Also physiologically important are the flavin-dependent monooxygenases. These systems can be divided into two groups. The first one consists in enzymes with flavin adenine dinucleotide (FAD) or flavin mononucleotide (FMN) as prosthetic groups, in which reduction of the flavin by reduced pyridine nucleotides, NADH or NADPH, and oxidation of substrates are carried out on a single polypeptide chain (3). Recently emerged a second class of enzymes which utilize the flavin, FAD or FMN, as a cosubstrate rather than a prosthetic group. A separate NAD(P)H:flavin oxidoreductase is thus required to supply reduced flavins that bind to the oxygenase component where they activate $O_2$ for substrate oxidation (Scheme 1) (4). Therefore, in this two-component system, NAD(P)H oxidation and the hydroxylation reaction are catalyzed by separate polypeptides. For years, the only prototype for this class of monooxygenases was the luciferase enzyme which has been the subject of numerous studies (3,5,6). However, it is only recently that the importance of such systems in oxidative metabolism, in particular within microorganisms, has been realized.

If we except the enzymes involved in the luciferase reaction, the flavin reductases associated with the oxygenase component display strong sequence homologies (4). These flavin reductases have been extensively studied from a biochemical (7-10) and a structural point of view (11,12). On the other hand, much less is known on the structure and the enzymatic mechanisms of the monooxygenase component.

The monooxygenase components can be further divided into two groups on the basis of the type of the flavin used in the reaction. The first group consists in $FADH_2$-utilizing monooxygenases which are represented by several well-characterized systems such as 4-hydroxyphenylacetate 3-monooxygenase (HpaB) of *Escherichia coli* (13) and 2,4,5-trichlorophenol monooxygenase from *Burkholderia cepacia* (14,15). 2,4,6-trichlorophenol monooxygenase (TcpA) from *Ralstonia eutropha* (16), phenol hydroxylase (Phe A1) from *Bacillus thermoglucosidasius* (9) and styrene monooxygenase (StyA) from *Pseudomonas flurescens ST* (17,18) belong to that family but have not been characterized yet. The second group consists in $FMNH_2$-utilizing monooxygenases which have been much less investigated in terms of their structure, mechanism and substrate specificity. Examples of such FMN-dependent systems are those involved in the synthesis of the antibiotic pristinamycin in *Streptomyces pristinaespiralis* (19,20), biodegradation of polyaminocarboxylates such as ethylenediamine tetracetic acid (EDTA) (21) or nitrilotriacetic acid (NTA) (22) by microorganisms, utilization of sulfur from aliphatic sulfonates (23) and desulfurization of fossil fuels by *Rhodococcus* species (24).

The results reported here concern the enzyme system we have selected to study the chemistry of FMN-utilizing monooxygenases. The system participates to the last step of the biosynthesis of the antibiotic actinorhodin in *Streptomyces coelicolor* (Scheme 2) which consists in the hydroxylation of the precursor dihydrokalafungin and the coupling of two molecules of the hydroxylated product (25-27).

Indications that this step was catalyzed by a two-component flavin-dependent monooxygenase came from genetic (25) and biochemical studies (7,10,25). Actinorhodin biosynthesis involves about 20 proteins whose corresponding genes are localized in the same region of the chromosome (27). Inactivation of the *actVB* gene resulted in the accumulation of dihydrokalafungin without any production of actinorhodin (25). Purification and characterization of ActVB demonstrated that ActVB was a NADH:FMN oxidoreductase, therefore suggesting that it serves to provide reduced FMN to a monooxygenase involved in dihydrokalafungin oxidation catalysis (7,10). The same phenotype was observed when a second region of the chromosome, named *actVA*, was inactivated (26) . This region contains 6 open reading frames (ORFs) (26). Here, we identify ActVA-ORF5 as the monooxygenase associated with ActVB. Overexpression in *E. coli* allowed us to purify it and to characterize the FMN-dependent activation of molecular oxygen by the ActVA-ActVB system.

## MATERIALS AND METHODS

*Materials.* FMN, NADH, DHAQ (1,5-dihydroanthraquinone) were purchased from Sigma or Aldrich. Other reagent-grade chemicals were obtained from Euromedex. Deazaflavin (5-Deaza-5-carbariboflavin) was a gift from Philippe Simon (Grenoble, France).

*Construction of ActVA-ORF5 expression plasmid*. Plasmid pIJ2305 containing the whole *actV* region (28) was isolated from *Streptomyces coelicolor* with the Plasmid Midiprep Kit (Qiagen). A*ctVA-ORF5* gene was amplified by PCR from pIJ2305 using the primer 5'-GGAATTCCATATGAGCGAGGACACGATGACC-3' (*EcoR1 and Nde1* restriction sites underlined) and the reverse primer 5'-CGCGGATCCTCAGCCATCGTTGCTCCTCCT-3' (*BamH1* restriction site underlined). The PCR product was digested by EcoR1-BamH1 and the resulting fragment was ligated into pUC18 (pUC18-ActVA). This plasmid was sequenced in order to confirm that no changes had been introduced in the *actVA-ORF5* sequence during PCR. The *Nde1-BamH1* fragment derived from pUC18-ActVA was subsequently cloned into pT7-7 at the same restriction sites, resulting in the plasmid named pActVA used for the overexpression of the *actVA-ORF5* gene product.

*Overexpression and purification of ActVA-ORF5. E. coli* BL21 (DE3) pLysS harboring pActVA was grown at 37 °C and 220 rpm in a 3-liter flask containing 1 liter of Luria and Bertani medium in the presence of 200 μg/ml ampicillin and 34 μg/ml chloramphenicol. Expression of ActVA-ORF5 was induced by adding IPTG to a final concentration of 380 μM when an $OD_{600nm}$ = 0.4 was reached. After 5-6 hours of growth, cells were harvested by centrifugation at 4 °C and suspended in 10 mM Tris-HCl pH 7.6, 10% glycerol (buffer A). All the following steps, except cell disruption, were performed at 4 °C. Extraction of soluble proteins was performed by lysozyme treatment (0.5 mg/ml lysozyme, incubation time 45 min at room temperature) followed by French Press disruption. The resulting cell extract was clarified by ultracentifugation at 45,000 rpm during 90 min (rotor TI 50.2 Beckman) and the supernatant was used as crude extract for purification. Crude extract (700 mg in proteins) was loaded onto a 60 ml Q Sepharose XL column (Pharmacia), previously equilibrated with buffer A. The column was then washed with 60 ml of buffer A plus 25 mM NaCl and the proteins were eluted with a linear 0-500 mM NaCl gradient. Fractions containing ActVA, eluted with 120 mM NaCl, were pooled and concentrated to 2 ml by ultrafiltration using a 10,000 MW regenerated cellulose membrane (Amicon). 30% of the total protein fraction was lost in this step, probably due to irreversible adsorption to the membrane. The concentrated fraction was loaded onto a Superdex 200 gel filtration column (120 ml, from Amersham Biosciences) previously equilibrated with buffer A and then eluted with the same buffer. Fractions containing pure ActVA, about 95 % as determined by SDS-PAGE electrophoresis, were pooled, concentrated as described above and stored at –80 °C.

*Preparation of apoActVB*. Overexprexion and purification of His-tagged ActVB was carried out as previously reported (10). All the following steps were performed at 4 °C. Purified His-tagged ActVB (2 mg) was loaded onto a 1 ml Ni-NTA column (Pharmacia) and treated with 1.5 column volumes of 2 M KBr and 2 M Urea, at acidic pH (29). Renaturation on the column was achieved with 4 column volumes of Tris-HCl 50 mM pH 7.6, 10 % glycerol and the protein was recovered with 3 column volumes of 500 mM imidazole, 25 mM Tris-HCl pH 7.6, 10% glycerol. As apoActVB was unstable in the presence of imidazole, the protein solution was immediately loaded onto a Nap 10 column and eluted with 25 mM Tris-HCl pH 7.6, glycerol 10%. The yield of the overall process was about 40%.

*Analytical Determinations*. The native molecular mass of ActVA was determined with an analytical Superdex 200 gel filtration (Amersham Biosciences) equilibrated with 10 mM Tris-HCl pH 7.6 containing 150 mM NaCl. Bovine serum albumin (66 kDa), ovalbumin (45 kDa), chymotrypsin (20.1 kDa) and cytochrome c (12.4 kDa) were used as markers of molecular mass. The void volume was determined with ferritin (450 kDa). Protein concentration was determined using the Bio-Rad protein assay reagent, with bovine serum albumin as a standard.

*Flavin Binding to ActVA-ORF5 and ActVB*. UV-Visible absorbance spectra were recorded with a Hewlett-Packard 8453 photodiode array instrument at 18°C. $K_d$ values of $FMN_{ox}$ for ActVB and ActVA were calculated from the variation of the absorbance values at the selected wavelengths, using eq 1:

$$(A_x - A_o)/(A_f - A_o) = ([FMN_{tot}] \times [protein_{free}])/(K_d + [protein_{free}]) \quad (1)$$

where $A_0$ is the initial absorbance, $A_x$ the absorbance after a given addition of the protein, $A_f$ the absorbance at the end of the titration. Making the hypothesis that ActVA and ActVB contain one binding site per monomer, $[FMN_{bound}]$ and $[protein_{free}]$ were calculated from the total concentrations of FMN ($[FMN_{tot}]$), and protein ($[protein_{tot}]$), using eq 2 and eq 3:

$$[FMN_{bound}] = (A_x - A_o)/(A_f - A_o) \times [FMN_{tot}] \quad (2)$$

$$[protein_{free}] = [protein_{tot}] - [FMN_{bound}] \quad (3)$$

ActVA fluorescence emission spectra were recorded at 18 °C with a Jasco FP 6500 spectrofluorimeter, using a 1 cm square cuvette (Helma). The excitation wavelength was set to 295 nm. The measured fluorescence spectra were corrected from inner filter effects due to $FMN_{ox}$ or $FMN_{red}$ absorption during excitation as well as during emission. The corrected fluorescence ($F_{corr}$) was calculated according to the method described in (30), using eq 4:

$$F_{corr} = F_{obs}\left[\sum_{x=1}^{x=10}\sum_{y=1}^{y=10}\left(10^{-\frac{\varepsilon ex.[FMN].x.L}{10}}\right)\left(10^{-\frac{\varepsilon em.[FMN].y.L}{10}}\right)\right] \qquad (4)$$

where L is the cuvette full path length (1 cm), [FMN] the total concentration of FMN, $\varepsilon_{ex}$, the extinction coefficient of FMN at excitation wavelength (295 nm), $\varepsilon_{em}$, the extinction coefficient of FMN for each emission wavelength between 300 and 400 nm. Since the inner filter effect is not homogenous in the entire cuvette, we divided it into 100 identical points and defined their position by x and y coordinates making the hypothesis that their surfaces are infinitely small. For each position and each wavelength between 295 and 400 nm, the factors representing the inner filter effect were calculated and were subsequently summed in order to determine an F corrected ($F_{corr}$) value. $F_{corr}$ precisely corresponds to the intrinsic protein fluorescence.

$K_d$ values were calculated from the variation of the corrected fluorescence intensity at 338 nm ($F_{corr}$) using eqs 1, 2 and 3.

*Anaerobic experiments*. Anaerobic experiments were performed in a Jacomex glove box equipped with a UV-visible cell coupled to a Uvikon XL spectrophotometer by optical fibers (Photonetics system). All the solutions were incubated anaerobically 2 hours before the beginning of each experiment. For FMN reduction, an FMN stock solution (500 µM) was anaerobically and quantitatively photoreduced by 30 min irradiation with a commercial slide projector placed at a distance of 3 cm, in the presence of 2 µM deazaflavin and 10 mM EDTA. For oxidation of reduced flavin, as a standard procedure, aerated water was prepared for each experiment by bubbling a 100 % oxygen gas for 30 min into an air-tight eppendorff tube containing 1 ml of water. $FMN_{red}$ oxidation was performed in the anaerobic glove box by injecting 5 μl of the oxygen saturated water (1 mM) into a 100 μL air-tight spectrophotometric cuvette with an Hamilton syringe.

## RESULTS

### *ActVA-ORF5 as the potential flavin-dependent hydroxylase*

Among the six genes present in the *actVA* region (26), only ORF5 displays sequence homologies with flavin-dependent hydroxylases. This becomes particularly obvious when the sequence of the corresponding protein is compared to the product of the *tdsC* gene from *Paenibacillus spA11-2* (31) and the product of the *pheA* gene (32) (Supplementary data Figure 1). TdsC is a FMN-utilizing monooxygenase associated with TdsD, a NADH:FMN oxidoreductase, and involved in the oxidation of dibenzothiophene (31). A similar system is found in *Rhodococcus erytropolis* with DszC acting as a FMN-utilizing monooxygenase and DszD as a NADH:FMN oxidoreductase displaying strong sequence homologies with ActVB (24). Therefore, this analysis suggests that *actVA-ORF5* gene product might be the monooxygenase associated with ActVB for the last step of actinorhodin biosynthesis.

### *Purification of ActVA-ORF5*

The details of the methods used to clone, express the *actVA-ORF5* gene in *E. coli* and purify the corresponding protein are given in the experimental section. Briefly, *E. coli* Bl 21 (DE3) pLysS strain was transformed with the expression vector pActVA-ORF5 derived from pT 7-7 containing the *actVA-ORF5* gene. Bacterial soluble extracts were subjected to an anion exchange chromatography and the enriched fractions, as judged by SDS-PAGE analysis, were further purified by gel filtration on a Superdex 200 column. The protein was obtained 95 % pure as determined by SDS-PAGE analysis (data not shown). No proteolysis occurred all through the purification process (data not shown). Determination of its N-terminal sequence (SEDTHT) confirmed that the purified protein was indeed the product of the *actVA-ORF5* gene. For the sake of simplicity, in the following, this protein is named ActVA. Analysis by mass spectroscopy showed that the protein, with a molecular mass of 39,715 Da, lacked the N terminal methionine. Chromatography of pure ActVA on a calibrated Superdex 200 gel filtration column showed that the protein was dimeric in solution, with an experimental mass of about 77,000 Da (data not shown). Finally, no evidence for a protein-bound chromophore could be obtained by UV-visible spectroscopy.

At room temperature the protein spontaneously precipitated in the presence of 250 mM NaCl. Thus, the following experiments using pure ActVA were carried out in 50 mM Tris-HCl pH 7.6 buffer, in the absence of salt.

*ActVA-ORF5, an active monooxygenase system*

The precursor of actinorhodin is dihydrokalafungin. However, this compound is not commercially available and difficult to obtain either from natural sources (25) or by chemical synthesis (33). Thus, we used 1,5-dihydroanthraquinone (DHAQ) as a substrate analog to check whether the ActVA protein was indeed able to catalyze an hydroxylation reaction in combination with ActVB. The assay mixture, under aerobic conditions, contained 200 µM NADH, 5 µM FMN, 50 µM DHAQ, 11 nM ActVB and various amounts of ActVA, in 2 ml of 50 mM Tris-HCl pH 7.6. After one hour at 30 °C, DHAQ and its oxidation products were extracted with MeOH and analyzed by ESI-MS. As shown in Figure 1, a peak at m/z = 255 was observed besides the peak at m/z = 239 corresponding to DHAQ. This proved the formation of a product corresponding to DHAQ plus one oxygen atom (+16), thus deriving from DHAQ by monooxygenation. This peak was absent from the mass spectrum when one of the following components of the system was excluded from the reaction mixture, FMN, NADH, ActVB or DHAQ (data not shown). In addition, a control reaction carried out within an anaerobic glove box did not result in the production of the oxygenated product either (data not shown). In conclusion, taken together, these preliminary experiments reported here clearly indicate that the product of the *actVA-ORF5* gene is a flavin-dependent monooxygenase.

*Interaction of FMN with ActVA and ActVB: dissociation constants*

In order to determine the mechanism of the transfer of reduced flavins between ActVB and ActVA, the dissociation constant ($K_d$) values for reduced and oxidized FMN with ActVB and ActVA proteins were determined.

As reported previously, a significant proportion (20-30 %) of the purified ActVB polypeptides contained one bound FMN (10). Removal of FMN for the production of ActVB in the pure apoprotein form, named apoActVB, was thus achieved first prior to the binding experiments. The use of the his-tagged form of ActVB bound to a Ni-NTA column allowed a fast and efficient procedure for the preparation of apoActVB, as described in the experimental section. The protein bound to the Ni-NTA column was treated first with a solution containing 2 M urea and 2 M KBr. This was followed by renaturation, elution with imidazole and desalting. The protein obtained from that process retained its ability to bind FMN, as judged from the appearance during incubation of apoActVB with FMN, of an absorption band, centered to 457 nm, characteristic of the ActVB:FMN complex (10). ApoActVB was shown to be active as a flavin reductase in a standard assay using NADH and riboflavin or FMN as substrates, with $K_m$ values for flavins (data not shown) comparable to those obtained with the as-isolated enzyme (10). Taken together, these data showed that apoActVB was fully active, well-folded and could thus be used in our following investigations.

Figure 2 shows the UV-visible spectrum of the oxidized form of FMN ($FMN_{ox}$), under three different conditions. The □ max value of the low-energy visible band is 445 nm for free $FMN_{ox}$, 438 nm for $FMN_{ox}$ in the presence of an excess of ActVA and 457 nm for $FMN_{ox}$ in the presence of an excess of ActVB. This showed that $FMN_{ox}$ bound to ActVB, as previously demonstrated (10), but also to ActVA. Therefore, the □ max value could be used as an indicator for whether $FMN_{ox}$ is free or bound to ActVA or ActVB.

As shown in Figure 3, aerobic titration of $FMN_{ox}$ with increasing amounts of ActVA, resulted in a shift of the $FMN_{ox}$ visible absorption band from 445 to 438 nm. An isobestic point was observed at 458 nm. A plot of bound FMN versus free ActVA, calculated from the fractional absorbance change at 440 nm as described in the experimental section, gave a $K_d$ value of 19.4 ± 6.3 □M for $FMN_{ox}$ with regard to ActVA (Inset of Figure 3).

A similar spectrophotometric titration of $FMN_{ox}$ with ActVB was performed. Again from the shift of the $FMN_{ox}$ absorption band, from 445 to 457 nm as the concentration of ActVB increased (isobestic point at 385 nm), a plot of bound FMN versus free ActVB, calculated from the fractional absorbance change at 480 nm, gave a $K_d$ value of 4.4 ± 0.6 □M for $FMN_{ox}$ with regard to ActVB (Supplementary data Figure 2).

Since $FMN_{red}$ does not exhibit marked absorbance bands, $K_d$ values for $FMN_{red}$ could not be obtained by the same spectrophotometric titration experiments. For that purpose, we took advantage of specific spectroscopic properties of $FMN_{red}$ as described in the following.

In the case of ActVB, binding of $FMN_{red}$ in its active site in the presence of $NAD^+$ results in the formation of a broad absorption band between 550 and 800 nm which is characteristic for a charge-transfer (CT) complex between $NAD^+$ and $FMN_{red}$ ((10) and Figure 4). Such a complex is also formed as the result of the oxidation of NADH by $FMN_{ox}$ in the active site of ActVB. As shown in Figure 4, the intensity of the charge transfer band increased upon successive addition of $FMN_{red}$ in the presence of an excess of $NAD^+$. A plot of the bound ActVB versus free $FMN_{red}$ determined from the variation of the absorbance of the charge transfer band at 680 nm gave a $K_d$ value of 6.6 ± 0.6 μM for $FMN_{red}$ with regard to the [ActVB-$NAD^+$] complex (Inset Figure 4).

Finally, in the case of ActVA, the well-established ability of $FMN_{red}$ to quench the fluorescence of the apoproteins (upon excitation of tryptophan residues at 295 nm) allowed us to design another method for the determination of the $K_d$ value for $FMN_{red}$. This approach could be also carried out with $FMN_{ox}$ allowing comparison with the results described above, obtained by UV-visible spectrophotometry. It should be mentioned that such experiments could not be carried out with ActVB since the protein is totally devoid of trytophan residues (27). Figure 5 shows a typical fluorescence quenching experiment for the ActVA-$FMN_{ox}$ combination. Addition of $FMN_{ox}$ led to a decrease of the fluorescence intensity of the ActVA apoprotein, excited at 295 nm, without any notable shift of the emission maximum (338 nm). Inner filters due to the absorption of FMN between 295 and 400 nm were corrected as described in the experimental section. As shown in the Inset of Figure 5, a plot of the bound ActVA versus free FMN determined from the fluorescence intensity at 339 nm gave a $K_d$ value of 26.3 ± 3.2 μM for $FMN_{ox}$ with regard to ActVA. This $K_d$ value was consistent with that obtained from the UV-visible titration experiments as described above. This indicates that the fluorescence quenching experiment for the determination of $K_d$ values is a reliable approach.

When a similar fluorescence quenching experiment was carried out in anaerobiosis (excitation at 295 nm), with ActVA, in the presence of increasing amounts $FMN_{red}$ produced by deazaflavin photoreduction, the $K_d$ value for $FMN_{red}$ was determined to be 0.39 ± 0.04 μM (Supplementary data Figure 3).

Table 1 summarizes the $K_d$ values obtained here for both $FMN_{ox}$ and $FMN_{red}$, characterizing their interaction with apoActVA and apoActVB. We deduce from these $K_d$ values that FMN transfer is thermodynamically favorable from ActVA to ActVB in the oxidized state and from ActVB to ActVA in the reduced state.

This conclusion was confirmed by experiments showing irreversible transfer of $FMN_{ox}$ from ActVA to ActVB and of $FMN_{red}$ in the reverse direction. In the first one, 10 μM of $FMN_{ox}$ was complexed to an excess of ActVA (140 μM) and, accordingly, the light absorption spectrum of the solution displayed a band centered at 438 nm (data not shown). Addition of increasing amounts of apoActVB resulted in a shift of that band to lower energies (data not shown). As shown in Figure 6, the □ max value increased to reach 457 nm after addition of about one equivalent of apoActVB with regard to ActVA, with no further changes for larger apoActVB/ActVA ratios. The final value of □ max was characteristic for the $FMN_{ox}$-ActVB complex, demonstrating that $FMN_{ox}$, initially in ActVA, had been transferred to and was bound to apoActVB. Taking into account the concentrations of the different components in the solution, a total transfer of $FMN_{ox}$ from ActVA to ActVB at a [ActVB]/[ActVA] ratio of 1 was consistent with the $K_d$ values reported in Table 1 [1]. In a second experiment, $FMN_{ox}$ was bound to an excess of ActVB (thus light-absorbing at 457 nm) and was reduced by NADH under anaerobic conditions (data not shown). Reduction could be monitored by visible spectroscopy not only from the disappearance of the 457 nm band, but also from the appearance of the broad absorption between 550 and 800 nm characteristic for the CT complex between $NAD^+$ and $FMN_{red}$ discussed above (Figure 4). Anaerobic addition of one equivalent of ActVA resulted in the instantaneous disappearance of about 90% of the latter absorption (data not shown). This indicated disruption of the charge-transfer complex, most likely resulting from a transfer of $FMN_{red}$ from ActVB to ActVA.

*Oxidation of $FMNH_2$ : evidence for a flavin-peroxide intermediate*

As shown above, $FMN_{red}$ is a good ligand for ActVA. In the following, the study of the reaction of the ActVA:$FMN_{red}$ complex with molecular oxygen is reported. In the experiment shown in Figure 7, 16 µM $FMN_{red}$ were anaerobically incubated with 140 µM ActVA, so that the flavin was totally complexed to the protein (Table 1). Addition of oxygen-saturated water (50 µM $O_2$ final concentration) resulted in the complete oxidation of the flavin within 5 minutes as shown by light absorption spectroscopy (Figure 7). The final spectrum with two bands at 370 nm and 438 nm, was superimposable to that of a preparation of $FMN_{ox}$–containing ActVA. Therefore, it showed that the flavin stayed in the protein during the whole oxidation reaction. In the inset of Figure 7, absorbances at 380 and 438 nm are reported as the function of time, after addition of $O_2$. At 380 nm, the kinetic trace indicated that the reaction with $O_2$ proceeded in two steps: an initial fast phase, occurring during mixing, followed by a second slower phase. At 438 nm, only one phase was observed, without evidence for the first rapid event observed at 380 nm. Without taking into account the initial fast phase at 380 nm, both kinetics at 380 and 438 nm were found exponential, with the same apparent first-order rate constant value $k = 0.44 \pm 0.02$ $min^{-1}$. Moreover, this rate constant was not a function of $O_2$ concentration (data not shown). This process was found to be much slower than the oxidation of free $FMN_{red}$ under same conditions (data not shown), suggesting that ActVA binds $FMN_{red}$ in an environment which protects it significantly from oxidation.

The spectrum in Figure 7 (and in supplementary data Figure 4) observed at the very beginning of the reaction (first spectrum after 15 s) is significantly different from that of $FMN_{ox}$ with a band at 380 nm more intense than at 440-450 nm. Together with the kinetic analysis (see above), this suggests a fast formation of an intermediate flavin species absorbing at 380 nm which then disappeared more slowly to generate $FMN_{ox}$. Based on the similarities of this first spectrum and those reported in enzyme systems involving flavin-peroxide species, usually absorbing at 380 nm (3), we might assign the observed intermediate spectrum to a FMN-OOH species resulting from the reaction between $FMN_{red}$ and $O_2$.

Oxidation of $FMN_{red}$ in complex with ActVA is also a very slow process when compared to that of $FMN_{red}$ in complex with ActVB. This was shown in the experiment described below. ActVB-$FMN_{ox}$ complex was treated anaerobically with a slight excess of NADH to generate ActVB-$FMN_{red}$. Oxidation of the flavin was initiated by the addition of aerated water and monitored by UV-visible spectroscopy. In Figure 8 are shown the initial oxidation rates Vi for the $O_2$-dependent oxidation of ActVB-$FMN_{red}$, in the presence of increasing amounts of ActVA. The observation that Vi decreased with increased ActVA was consistent with the notion that $FMN_{red}$ spontaneously moved from ActVB to ActVA and was oxidized more slowly within the latter than within ActVB. As shown in Figure 8, in the presence of one equivalent of ActVA compared to ActVB, the oxidation rate was similar to that of ActVA-$FMN_{red}$ alone. This is in agreement with the much better affinity of $FMN_{red}$ for ActVA than for ActVB. In addition, formation of $FMN_{ox}$ within ActVB occurred with no observable intermediate species absorbing at 380 nm (data not shown). Therefore a slow oxidation process via a FMN-OOH species is specific for ActVA.

The supplementary data Figure 4 shows one of those experiments of Figure 8, in which one equimolar amount of ActVA with respect to ActVB was added in the reaction mixture. The difference with the experiment shown in Figure 7 is that the $FMN_{red}$ is provided in the form of an ActVB-$FMN_{red}$ complex to the aerated solution of ActVA. The results shown in supplementary data Figure 4 are very similar to that of Figure 7, with two observable reaction steps and a comparable rate constant for the second slow step. Again, 15 s after $O_2$ addition, the UV-visible spectrum was different from that of $FMN_{ox}$, with a single band at 380 nm, in agreement with a FMN-OOH intermediate within ActVA. This result suggests that also in that case, $FMN_{red}$ oxidation occurred in ActVA, as a consequence of a fast transfer of $FMN_{red}$ from ActVB to ActVA. Furthermore it shows that the oxidation reaction is not affected by the presence of ActVB. The only noticeable difference with the experiment in absence of ActVB (Figure 7) was that the final spectrum of the product $FMN_{ox}$ exhibited a maximum at 457 nm showing that at the end of the reaction, $FMN_{ox}$ was bound to ActVB.

*Interaction of ActVA with ActVB*

In order to verify that ActVA and ActVB might not form a complex in solution, the following experiment has been carried out. When an equimolar mixture of His-tagged ActVB and ActVA was loaded onto the Ni-NTA column, ActVA was totally recovered in the run-through fraction and could not be detected in the imidazole fraction containing ActVB (data not shown). These data suggest that there is no association between ActVA and ActVB.

## DISCUSSION

The two-component flavin-dependent monoxygenases seem to be widely distributed within the bacterial world and used in a large variety of biosynthetic and metabolic reactions. However, while the flavin reductase component of these systems, which provides reduced flavins required for $O_2$ activation, has been the subject of detailed mechanistic and structural studies, very little is known on the monooxygenase component, where the oxidation takes place.

Our investigation of the final step of actinorhodin synthesis, which consists in the conversion of dihydrokalafungin into actinorhodin (Scheme 2) led us to characterize such a two-component FMN-dependent monooxygenase system. We first reported the purification and the characterization of ActVB, the NADH:FMN oxidoreductase associated with the reaction (10). The oxidation reaction itself is not a trivial reaction since it involves two oxidation steps, a dimerization and an aromatic hydroxylation. Not only is the order of the two steps still unknown but it is also not sure whether there are different genes for each step.

The ActVA region of the *act* cluster carries several genes involved in actinorhodin biosynthesis (26). Because of the general difficulties in purifying the enzymes of that region coupled with the lack of commercial substrates required for biochemical studies, so far only ORF6 was assigned to a defined function, catalysis of the conversion of 6-deoxydihydrokalafungin to dihydrokalafungin (34,35). However, we speculated that the ORF5 product, named ActVA here for sake of simplicity, could be the enzyme partner of ActVB for catalyzing one or the two steps of the oxidation of dihydrokalafungin to actinorhodin for the following reasons. First, a DNA fragment containing the *actVA-ORF5* region could cause *Streptomyces* sp. AM 7161, the producer of medermycin, an analog of actinorhodin which lacks the C-8 hydroxyl group, to make mederrhodin A, which is hydroxylated at C-8 (36). Secondly, as shown here, ORF5 displays significant similarities with genes encoding flavin-dependent monooxygenases (Supplementary material Figure 1). Our results strongly support this hypothesis. With a pure preparation of ActVA, reported here for the first time, we clearly show that the ActVA-ActVB combination can catalyze an hydroxylation reaction, using 1,5-dihydroanthraquinone as a substrate analog. Oxidation absolutely requires NADH, FMN and molecular oxygen, supporting the notion that ActVB serves for the production of reduced flavins whereas ActVA uses the latter to activate oxygen and oxidize the organic substrate. Accordingly, even though it was purified with no chromophore, we have shown that ActVA is able to bind $FMN_{ox}$, which displays a specific light absorption spectrum with a low-energy transition at 438 nm. ActVA is also able to bind $FMN_{red}$, to which it provides protection from oxidation by air. With the true substrate, dihydrokalafungin, when it is available, this in vitro system will allow us to conclude whether dimerization and hydroxylation is catalyzed by the same enzyme and, if not, which step is occurring first.

An interesting issue concerns the transfer of reduced flavin from one protein (ActVB) to another (ActVA), its control and whether specific interactions between ActVB and ActVA are required for that purpose.

The $K_d$ values for FMN, both in the reduced and the oxidized form, characterizing their interaction with each component, ActVA and ActVB, have been carefully determined. This is the first time that all these values are given with this class of enzymes. They indicate that the flavin transfer between ActVA and ActVB is under a thermodynamic control, with a preference of $FMN_{ox}$ for ActVB and a much stronger affinity of ActVA for $FMN_{red}$ ($K_d = 0.39$ μM). The data showed that $FMN_{ox}$ can be transferred from ActVA to ActVB (Figure 6) and also that $FMN_{red}$ generated in ActVB by reduction of $FMN_{ox}$ either by NADH or by photoreduction can be removed from ActVB and transferred to ActVA. This was also in good agreement with the observation that the oxidation of ActVB-$FMN_{red}$ by oxygen is greatly slowed down upon addition of increasing amounts of apoActVA, as a consequence of $FMN_{red}$ transfer to ActVA where oxidation occurs slowly (Figure 8).

All these results can thus be fitted into a very simple mechanism shown in Scheme 3. At the very beginning of the reaction, $FMN_{ox}$ binds to ActVB where it is reduced by NADH. The resulting $FMN_{red}$ then diffuses out of ActVB and tightly binds to ActVA where it reacts with oxygen. In this step a reactive intermediate is formed and used to hydroxylate the substrate thus generating $FMN_{ox}$ as will be discussed below. In the absence of the substrate it is likely that the intermediate decomposes to form hydrogen peroxide. Then, in all cases, $FMN_{ox}$ diffuses out of ActVA and ends up into ActVB ready to start a new cycle. In this mechanism, based on a thermodynamic control of the reaction, we do not find it necessary to invoke a specific interaction between the two proteins and a complex channeling mechanism allowing FMN to travel from one protein to another without equilibrating with FMN free in solution. In fact, we did not get any evidence for a complex between ActVB and ActVA. It is actually the case for most two-component flavin-dependent monooxygenase systems that the flavin reductase component and the monooxygenase component do not form a complex. Furthermore, in most cases, the oxidoreductase component can be replaced by many flavin reductases, including the non-homologous ones from other microorganisms. Therefore direct protein-protein interactions are in general not essential.

In a few cases it has been reported that the monooxygenase component stabilizes C-4a-flavin hydroperoxides in the absence of the substrate (3). One of the characteristics of these species, which derive from the reaction of reduced flavins with molecular oxygen, is their light absorption spectrum with a broad unique band at around 360-390 nm and the absence of the absorption band at 440-450 nm present in the spectrum of oxidized flavins. In the present study during oxidation of ActVA-$FMN_{red}$ or oxidation of ActVB-$FMN_{red}$ in the presence of ActVA, we have observed that the reaction, in the absence of substrate, occurs in two steps. In the first one, a species absorbing at 380 nm is rapidly generated, and in the second one, the latter is slowly converted to $FMN_{ox}$. We thus assign the 380 nm-absorbing species to a flavin hydroperoxide, FMN-OOH. The accumulation of the observed peroxide and its slow decomposition in ActVA is consistent with the protective effect of ActVA with regard to flavin oxidation by oxygen as discussed above. It is very likely that the flavin hydroperoxide is the oxidizing agent which attacks the substrate within ActVA. Further studies will address the substrate specificity of this interesting enzyme system as well as the chemistry of the reaction between the flavin hydroperoxide and the substrate.

ACKNOWLEDGEMENTS

We acknowledge Yves Dupont, Vincent Forge and Elizabeth Mintz for fruitful discussions. We thanks Domenica Gräfin Von Der Schulenburg for reading the manuscript.

FOOTNOTES

[1] In a presence of both ActVB and ActVA, because FMN is in binding equilibrium with both proteins, the concentration of bound FMN can be calculated with the following equation: $[\mathrm{FMN}]_{bound} = ([\mathrm{FMN}]_{tot}\ [\mathrm{ActVB}]_{free})/[(K_{d\mathrm{ActVB}}(1+([\mathrm{ActVA}]_{free})/(K_{d\mathrm{ActVA}}))+[\mathrm{ActVB}]_{free})]$

FIGURE LEGENDS

**Figure 1**

ActVA-ActVB hydroxylase activity. ActVB (11 nM) and ActVA (5 µM) were aerobically incubated with $FMN_{ox}$ (5 µM), NADH (200 µM) and DHAQ (1,5-dihydroanthraquinone) (50 µM), in 50 mM Tris-HCl pH 7.6, during one hour at 30 °C. Products were then extracted with methanol and analyzed by ES-MS. Are shown the mass spectrum of the organic extract and the structure of DHAQ (m/z = 239).

**Figure 2**

Light absorption spectra of $FMN_{ox}$. 10 µM $FMN_{ox}$ was dissolved in 10 mM Tris-HCl pH 7.6, alone (□), in the presence of 200 µM ActVA (○), or 40 µM ActVB (◆).

**Figure 3**

Spectrophotometric titration of $FMN_{ox}$ with ActVA. 10 µM $FMN_{ox}$ in 50 mM Tris-HCl pH 7.6 was titrated with ActVA. From the bottom to the top, the concentrations of protein used were 0, 12, 34, 72 and 91 µM. In the Inset is shown a plot of bound $FMN_{ox}$ versus concentration of free ActVA, determined from the fractional absorbance change at 440 nm. A $K_d$ value of 19.4 ± 6.3 µM was determined.

**Figure 4**

Spectrophotometric titration of ActVB with $FMN_{red}$ in the presence $NAD^+$. ApoActVB (24 µM) was incubated in anaerobiosis with increasing amounts of $FMN_{red}$ in the presence of 2 mM $NAD^+$ in 50 mM Tris-HCl pH 7.6 at 18 °C. From the bottom to the top are shown the 500-800 nm absorption spectra after addition of 0, 10, 18, 37, 53 and 106 µM of $FMN_{red}$. In the Inset is shown a plot of bound ActVB as a function of free $FMN_{red}$, determined from the fractional absorbance change at 680 nm. A $K_d$ value of 6.6 ± 0.6 μM was determined.

**Figure 5**

Fluorimetric titration of ActVA with $FMN_{ox}$. ActVA (37 μM) was titrated by different concentration of $FMN_{ox}$ in 50 mM Tris-HCl pH 7.6. From the top to the bottom are shown the ActVA fluorescence spectra (excitation at 295 nm) after addition of 0, 12, 36, 85 and 360 μM of $FMN_{ox}$. Inner filter effects due to the absorption of $FMN_{ox}$ at 295 nm and between 300-420 nm were corrected (see Materials and Methods). In the Inset is shown a plot of bound ActVA as a function of free $FMN_{ox}$, determined from the fluorescence intensity at 339 nm. A $K_d$ value of 26.3 ± 3.2 μM was determined.

**Figure 6**

$FMN_{ox}$ transfer from ActVA to ActVB. ActVA (140 μM) was mixed with $FMN_{ox}$ (10 μM) in 50 mM Tris–HCl pH 7.6. Increasing amounts of apoActVB were added and the maximal absorption wavelengths were determined and plotted as a function of the ActVB/ActVA ratio.

**Figure 7**

Reaction of ActVA:$FMN_{red}$ complex with molecular oxygen. $FMN_{red}$ (16 μM) was incubated anaerobically with an excess of ActVA (140 μM) in 10 mM Tris-HCl pH 7.6. The reaction was initiated by the addition of 50 μM $O_2$. From the bottom to the top, are shown the spectra recorded at 15 s, 82 s, 198 s, and 740 s. Inset shows the variation of absorbance at 380 nm (□) and 438 nm (O) as a function of time. Lines were calculated for best fit to a exponential model.

**Figure 8**

Effect of ActVA on the initial velocity of ActVB:$FMN_{red}$ complex oxidation. $FMN_{ox}$ (16 µM) in the presence of apoActVB (50 µM) was reduced with one equivalent of NADH, in 10 mM Tris-HCl pH 7.6. Then, the mixture was anaerobically incubated with different amounts of ActVA. $O_2$ (50 µM final concentration) was rapidly added and the initial velocity of $FMN_{red}$ oxidation was calculated from the increase of the absorbance at 457 nm as a function of time.

Table 1: Dissociation Constants for $FMN_{ox}$ and $FMN_{red}$ complexes with ActVA and ActVB, determined in 50 mM Tris-HCl pH 7.6, at 18 °C.

| | ActVA | ActVB |
|---|---|---|
| $FMN_{ox}$ | 26.3 ± 3.2[a] μM<br>19.4 ± 6.3[b] μM | 4.4 ± 0.6[b] μM |
| $FMN_{red}$ | 0.39 ± 0.04[a] μM | 6.6 ± 0.6[b, c] μM |

[a] values obtained by fluorimetric titration

[b] values obtained by UV-visible absorption titration

[c] in the presence of an excess of $NAD^+$

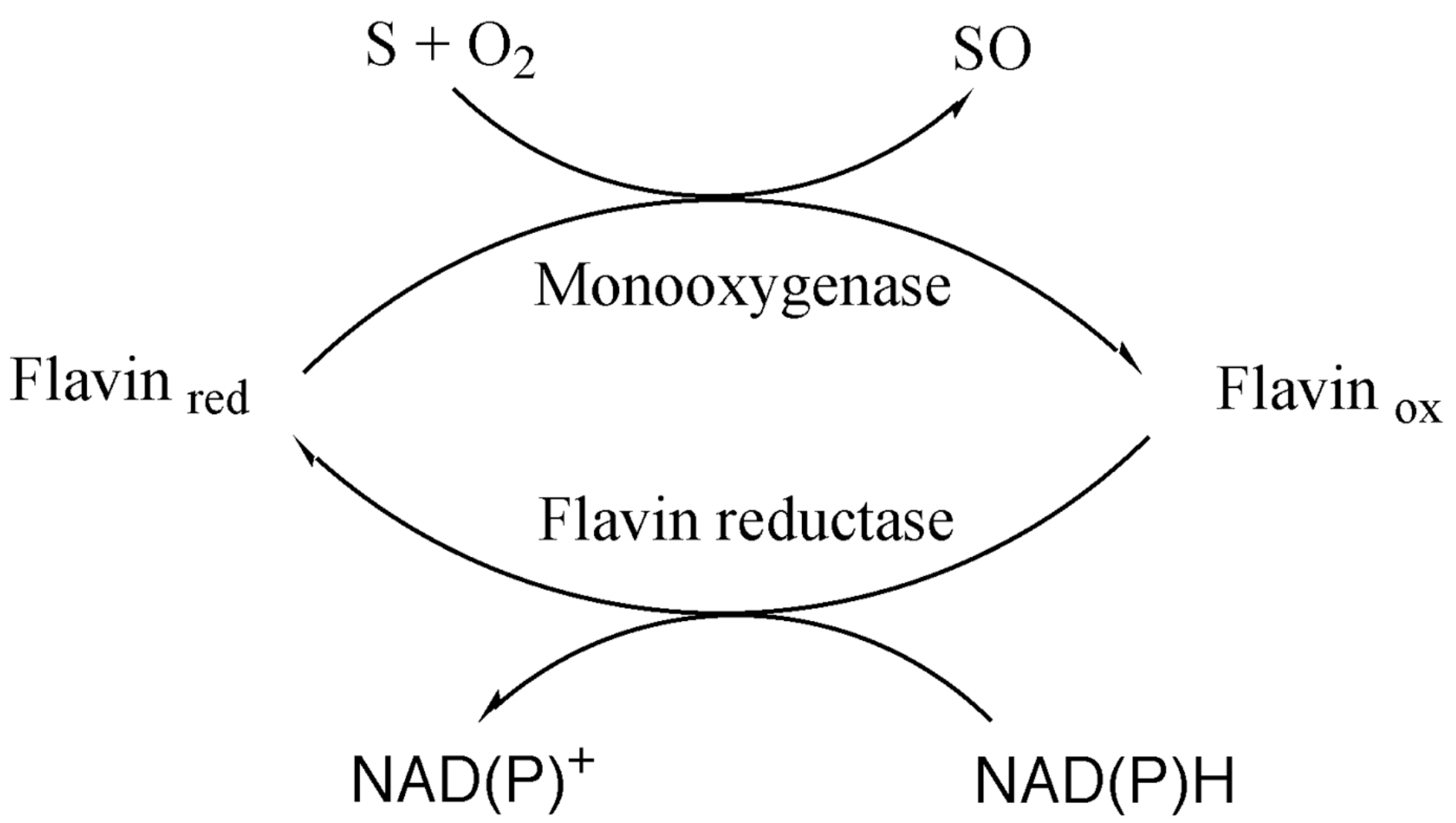

Scheme 1

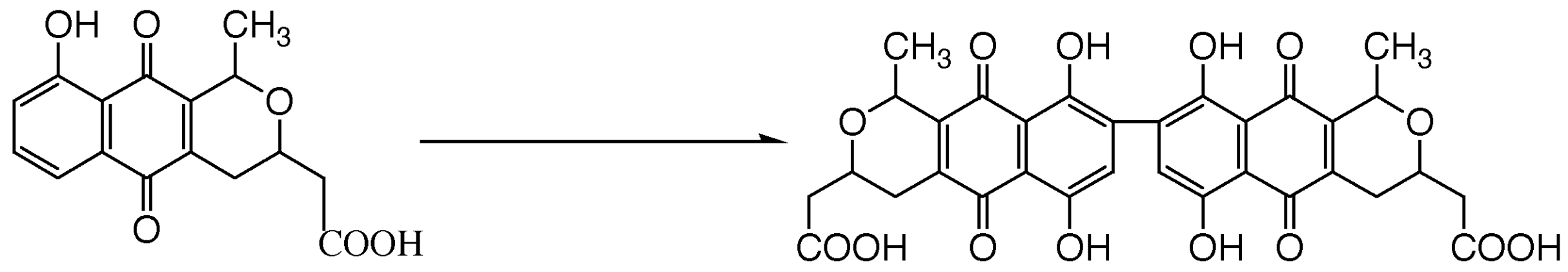

Scheme 2

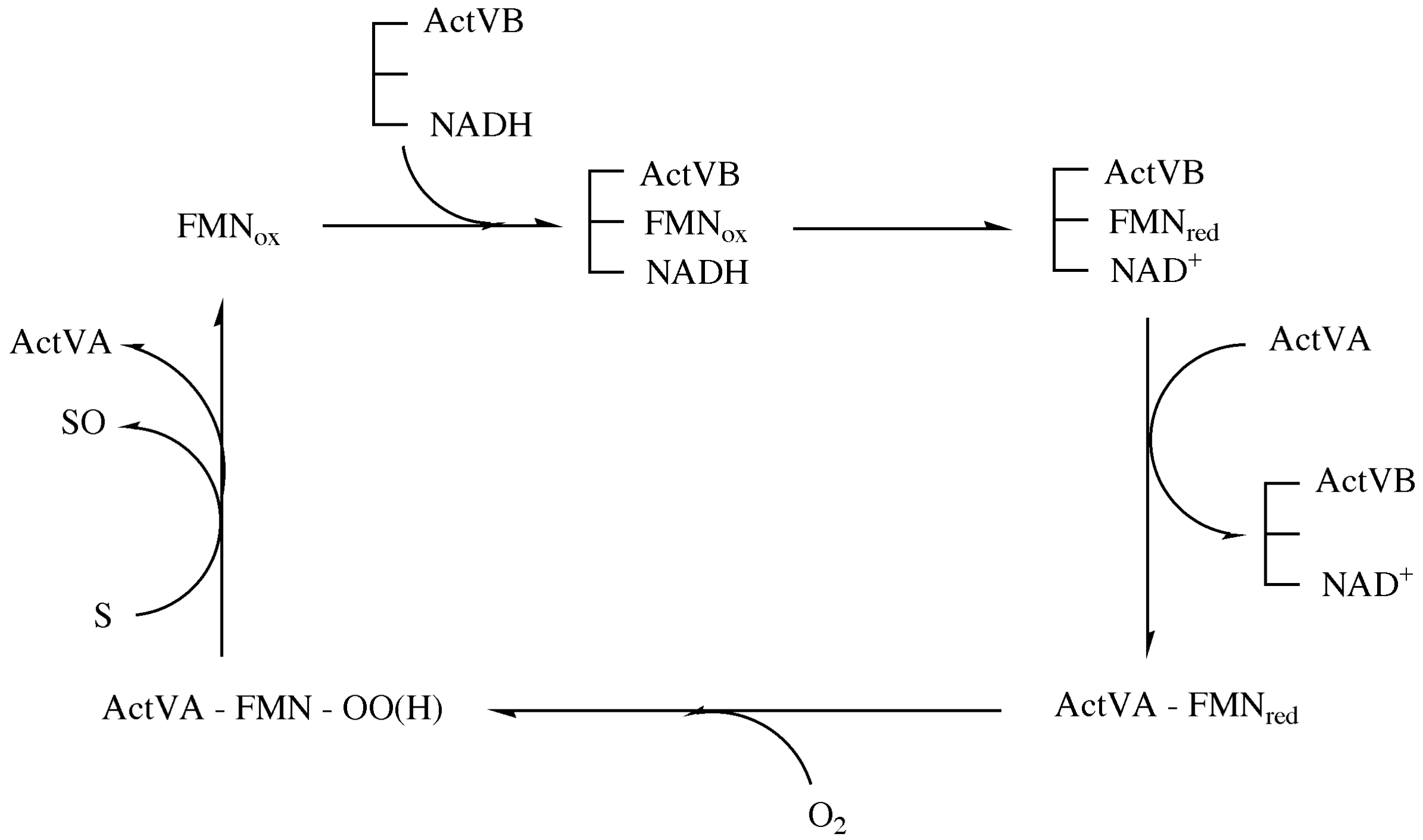

Scheme 3

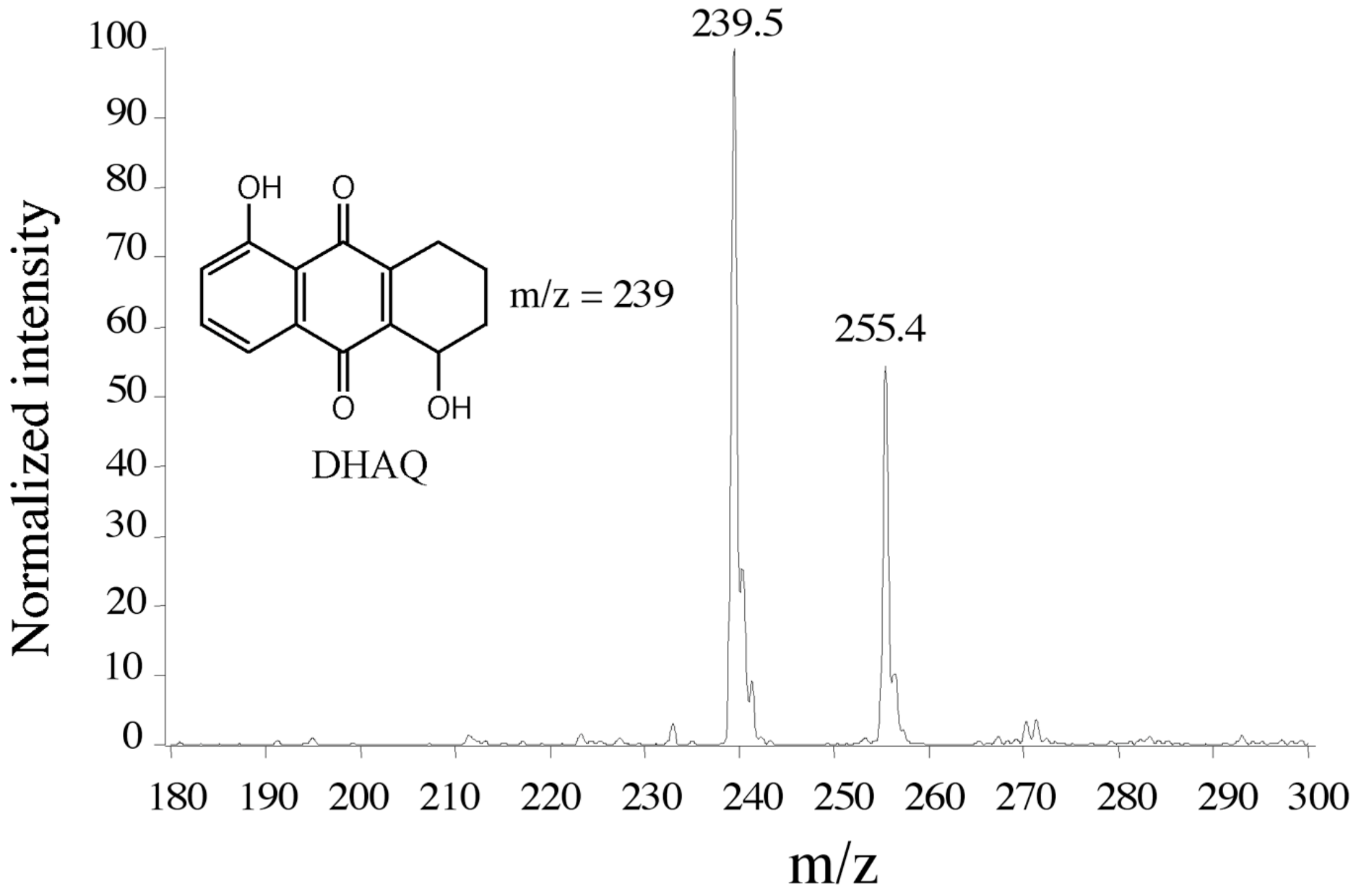

Figure 1

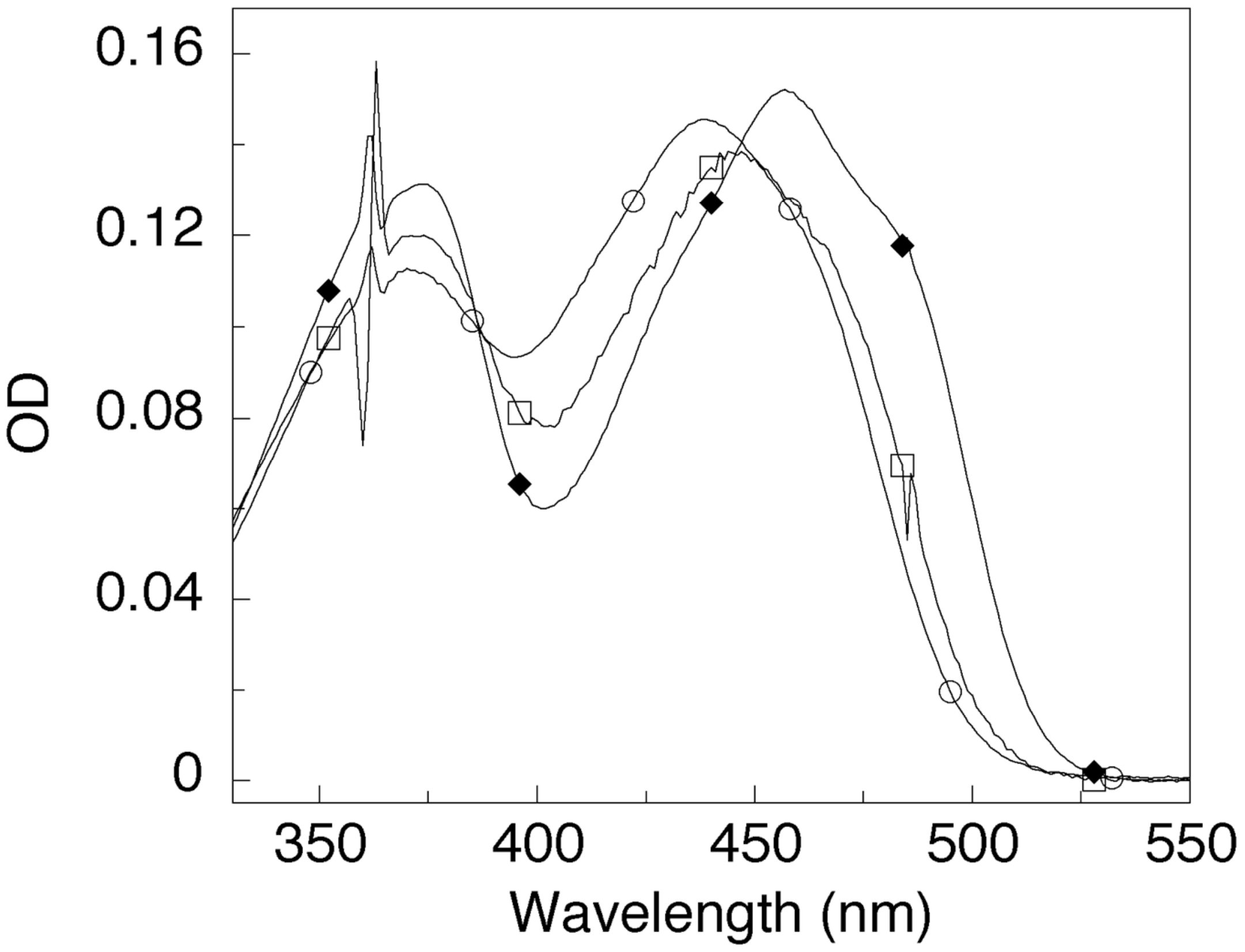

Figure 2

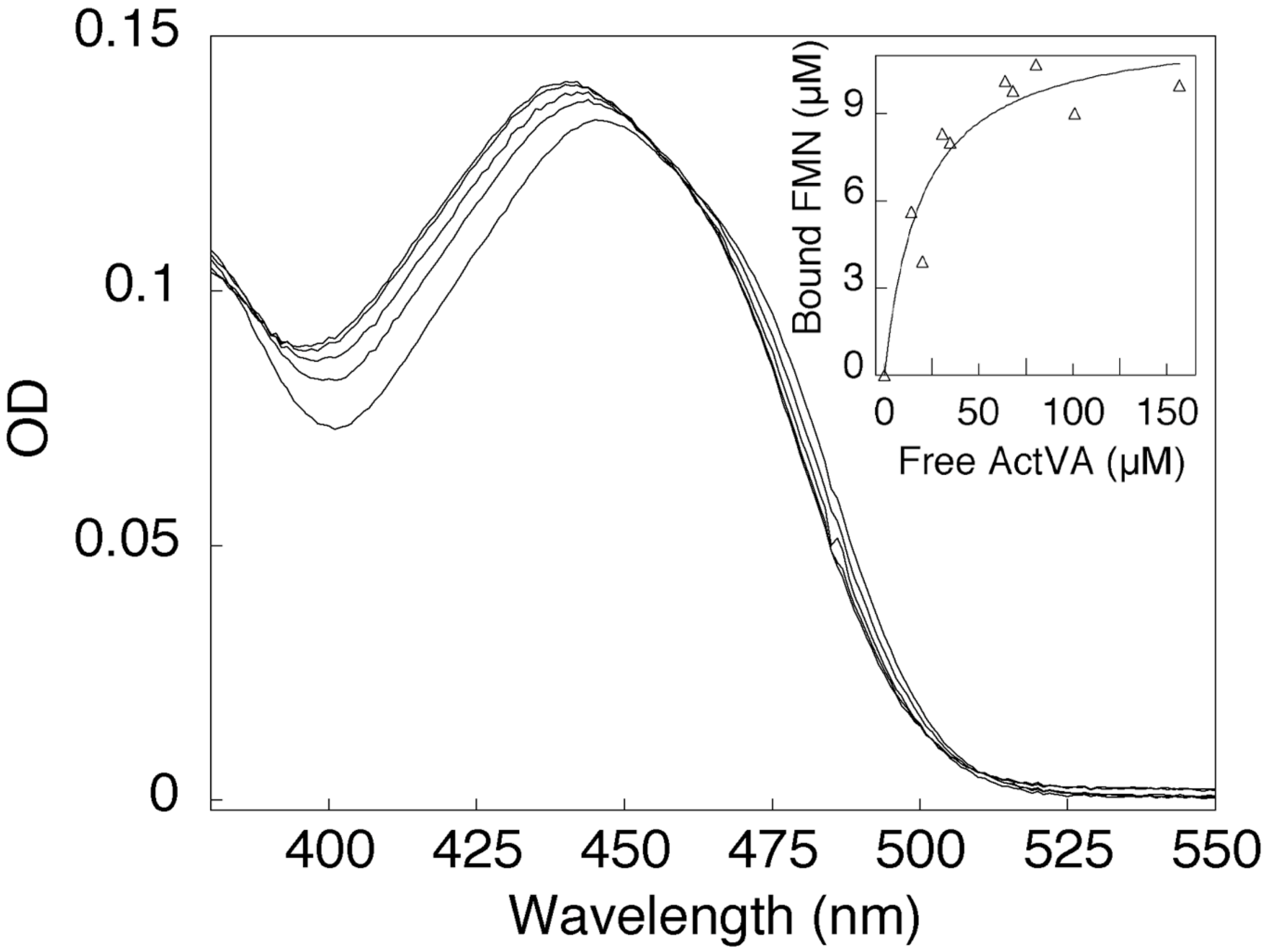

Figure 3

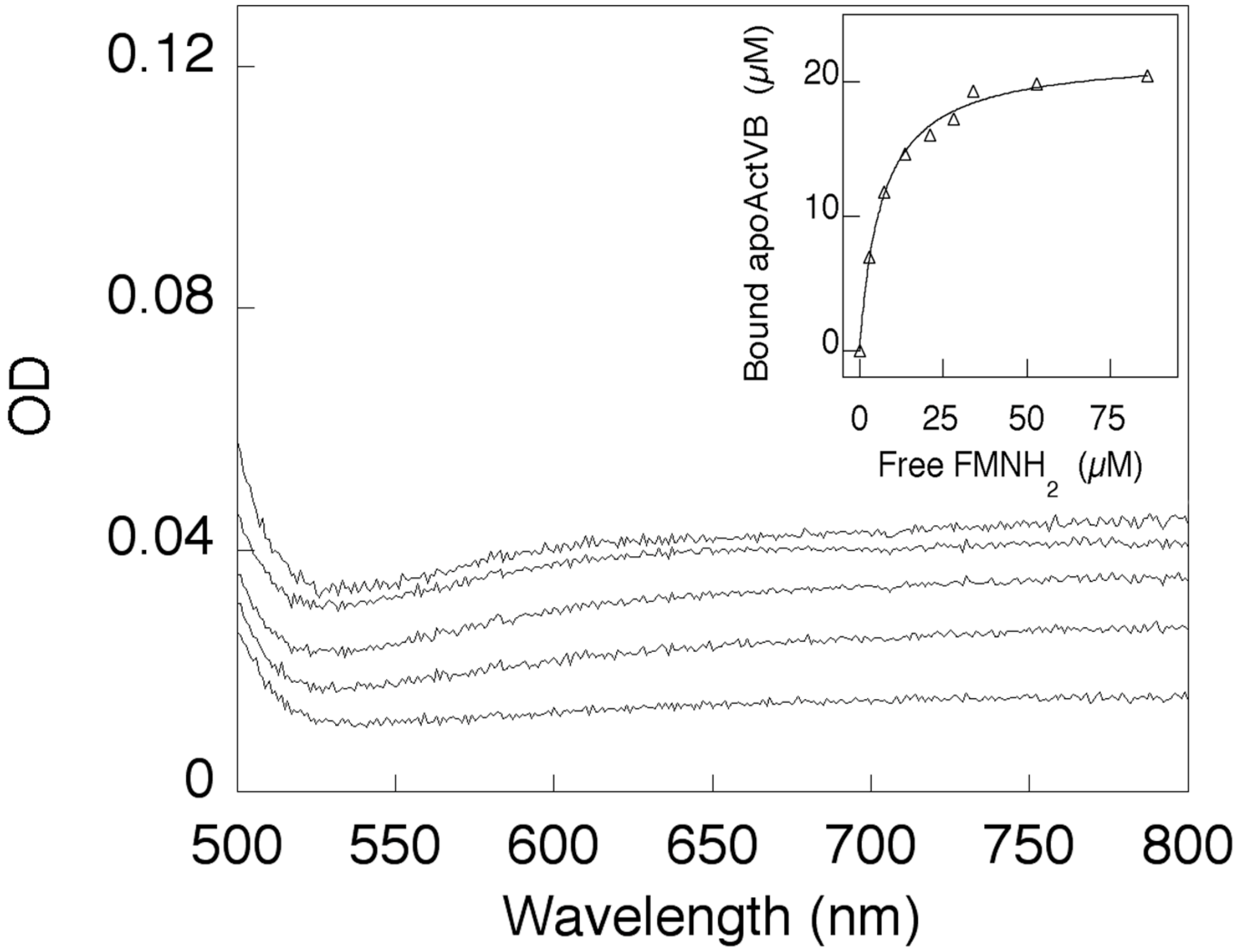

Figure 4

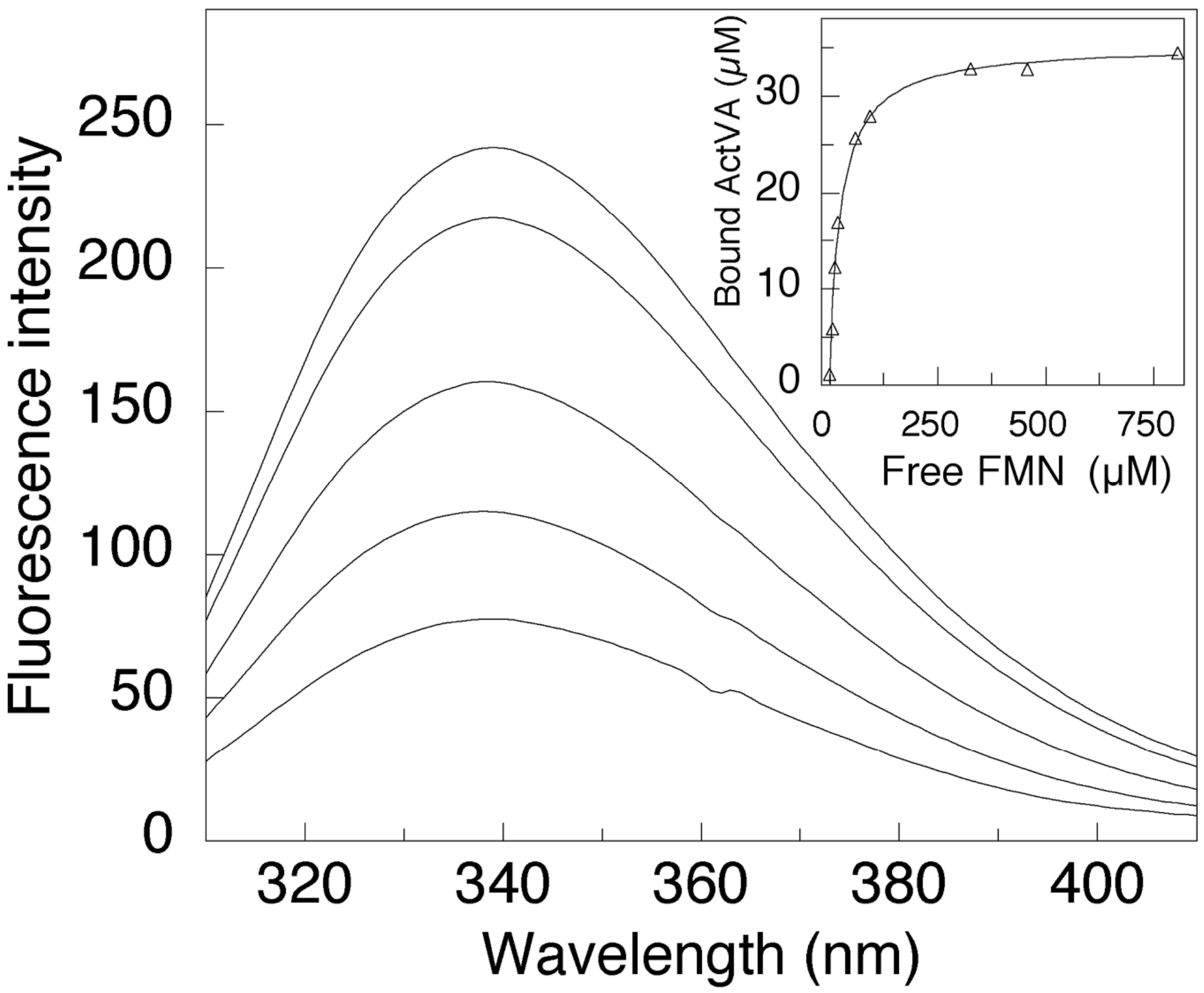

Figure 5

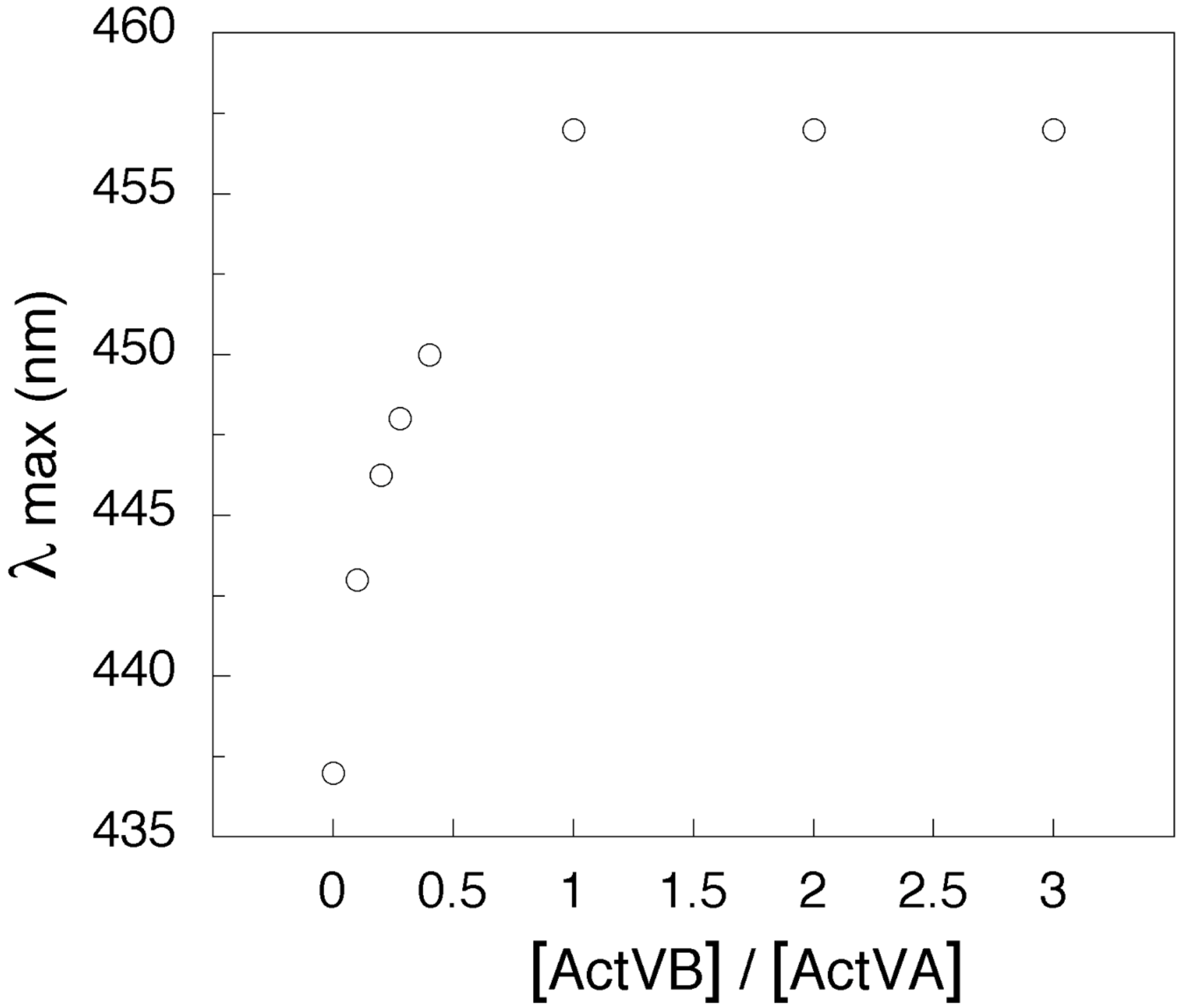

Figure 6

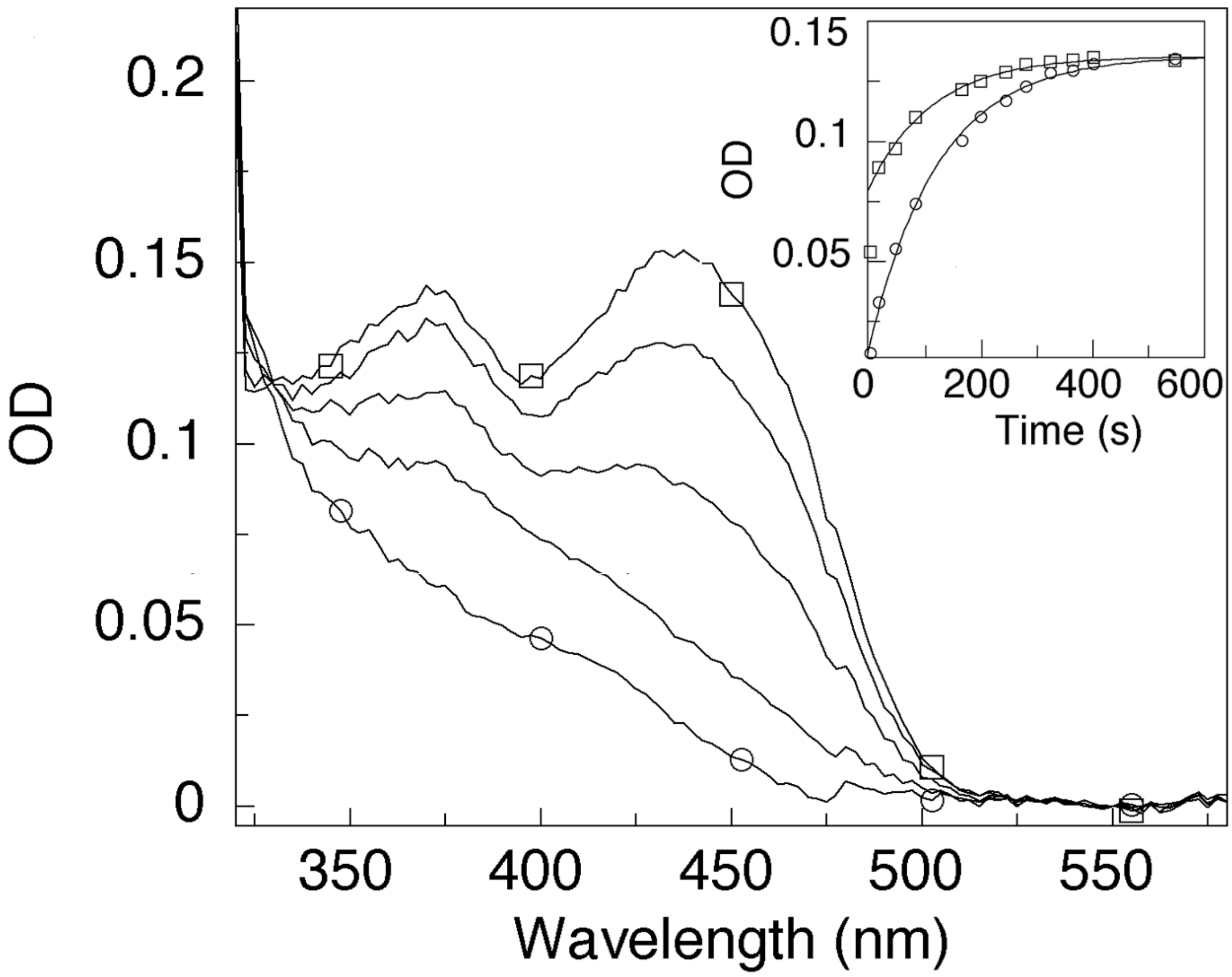

Figure 7

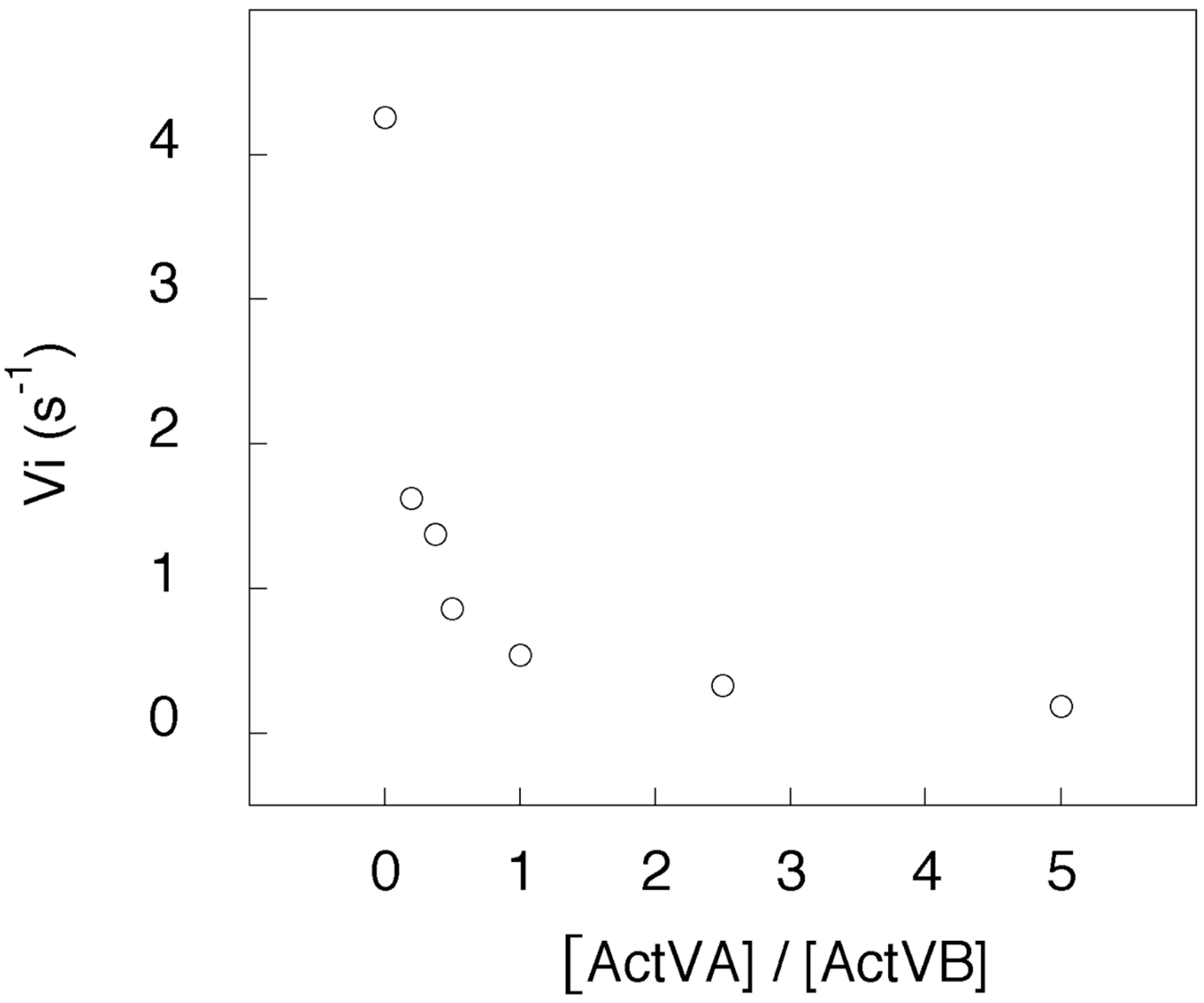

Figure 8

Supplementary data Figure 1

```
ActVA MSEDTMTQER----PSLTAHARRIAELAGKRAADAEQQRRLSPDVVDA 44
PheA  MEKNKMLIEEKLDTAALLAKAEEIGRIAEEEAGEADRNACFSDRVARA 48
TdsC  MRTIHANSSAVREDHRALDVATELAKTFRVTVRERERAGGTPKAERDA 48
      *       .           * .:..     . : ::    .     *

ActVA VLRAGFAAHFVPVAHGGRAATFGELVEPVAVLGEACASTA-------- 84
PheA  IKEAGFHKLMRPKQYGGLQVDLRTYGEIVRTVARYSVAAG-------- 88
TdsC  IRRSGLLTLLISKERGGLGESWPTVYEAIAEIASADASLGHLFGYHFS 96
      : .:*:   : .   **         * :  :.   .: .

ActVA --WYASLTASLGRMAAYLPDEGQAELWSDGPDALIVGALMPLGRAEK- 129
PheA  --WLTYFYSMHEVWAAYLPPKGREEIFGQG--GLLADVVAPVGRVEK- 131
TdsC  NFAYVDLFASPEQKARWYPQAVRERWFLGNASSENNAHVLDWRVTATP 144
          . : :     * : *   : . :  .  .     :     . .

ActVA -TPGGWHVSGTWPFVSVVDHSDWALICAKVGE------EPWFFAVP-- 168
PheA  -DGDGYRLYGQWNFCSGVLHSDWIGLGAMMELPDGNSPEYCLLVLP-- 176
TdsC  LPDGSYEINGTKAFCSGSADADRLLVFAVTSRDPNGDGRIVAALIPSD 192
         ..:.: *   * *   .:*   : *          .     :*

ActVA RQEYGIVDSWYPMGMRGTGSNTLVLDGVFVPDARACTRAAIAAG---L 213
PheA  KSDVQIVENWDTMGLRASGSNGVLVEGAYVPLHRIFPAGRVMAHGKPV 224
TdsC  RAGVQVNGDWDSLGMRQTDSGSVTFSGVVVYPDELLGTPGQVTDAFAS 240
      :    :  .* .:*:* :.*. : ..*. *   .        :

ActVA G--PDAEAICHTVPMRAVNGLAFALPMLGAARGAAAVWTSWTAGRLAG 259
PheA  GGDYDENDPVYRMPFMPLFLLGFPLVSLGGDERLVSLFQERTEKRIRV 272
TdsC  GSKPSLWTPITQLIFTHLYLGIARGALEEAAHYSRSHSRPFTLAGVEK 288
      *   .       : :  :           . .   :     *   :

ActVA PTGQNAVSSQDRVVYEHTLARATGEIDAAQLLLERVAAVADAGSATG- 306
PheA  FKGG--AKEKDSAASQRLLAEMKTELNAMEGIVEQYIRQLEACQKEGK 318
TdsC  ATEDPYVLAIYGEFAAQLQVAEAGAREVALRVQELWERNHVTPEQRGQ 336
       .    .         :  .      :.   : *       : .  *

ActVA VLVGRGARD-----CALAAELLTAATDRLFASAGTRAQAQDSP-MQRL 348
PheA  TVMNDMEREQLFAWRGYVAKASANIAVRTLLTLGGNSIFKGDP-VELF 365
TdsC  LMVQVASAK------IVATRLVIELTSRLYEAMGARAAASRQFGFDRF 378
       ::     .        .:.     : *   : * .:  . .  .: :

ActVA WRDVHAAGSHIGLQFGPGAALYAGELLRRSNDG---             381
PheA  TRDLLAVAAHPNSLWEDAMAAYGRTIFGLPGDPVW-             400
TdsC  WRDARTHTLHDPVAYKIREVGNWFLNHRFPTPSFYS             414
       **  :   *    :    .         .
```

Sequence alignment between ActVA from *Streptomyces coelicolor*, PheA from *Bacillus stearothermophilus* and TdsC from *Paenibacillus sp. A11-2*. Identical amino acids are shown in gray.

Supplementary data Figure 2

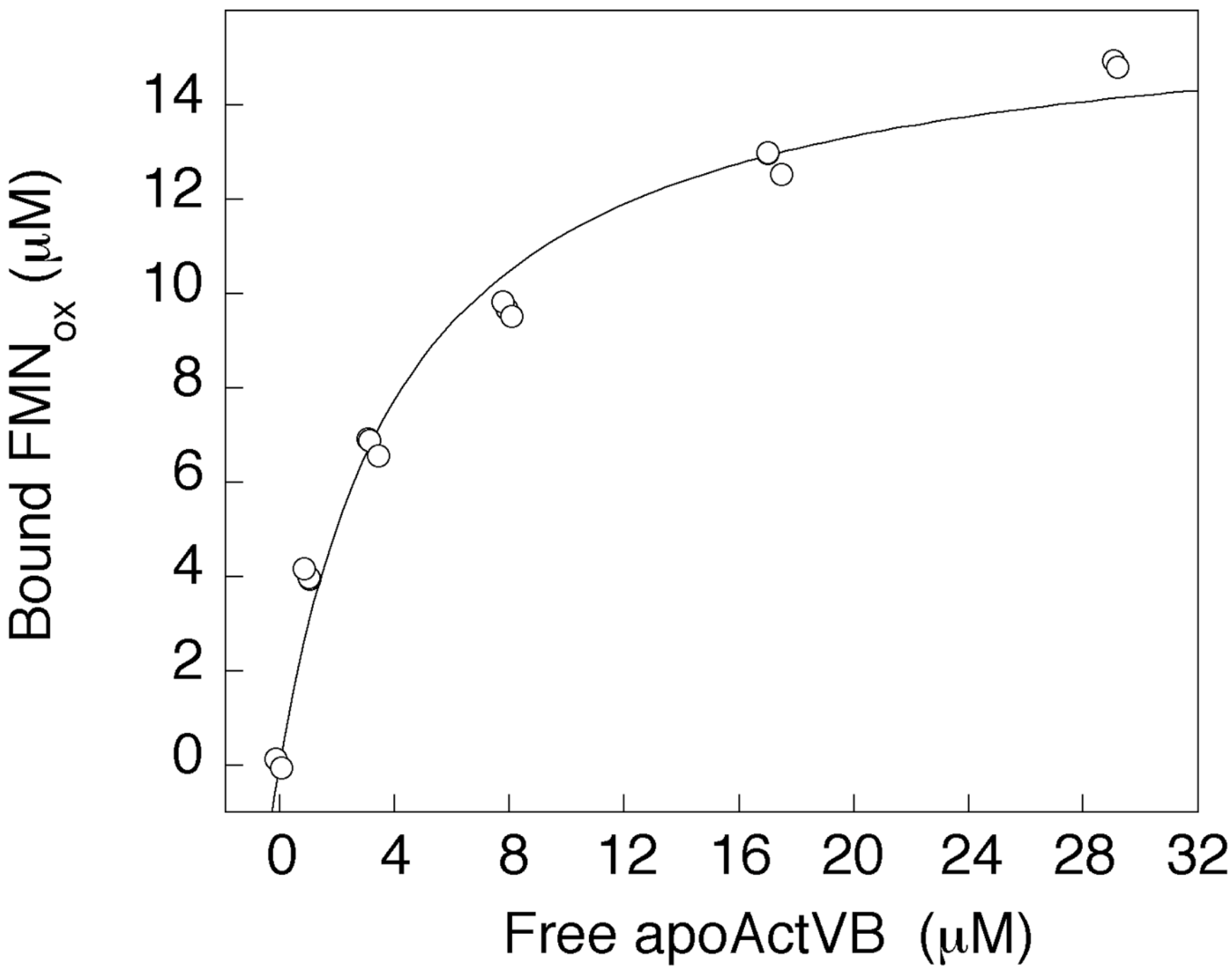

Determination of the $K_d$ value for $FMN_{ox}$ with regard to apoActVB. $FMN_{ox}$ (10 μM) was incubated with increasing amounts of apoActVB in Tris-HCl 50 mM pH7.6 at 18 °C. UV-visible spectra were recorded and the absorbance variation at 480 nm was used to determine the amount of bound $FMN_{ox}$ per apoActVB added as discussed in experimental section.

Supplementary data Figure 3

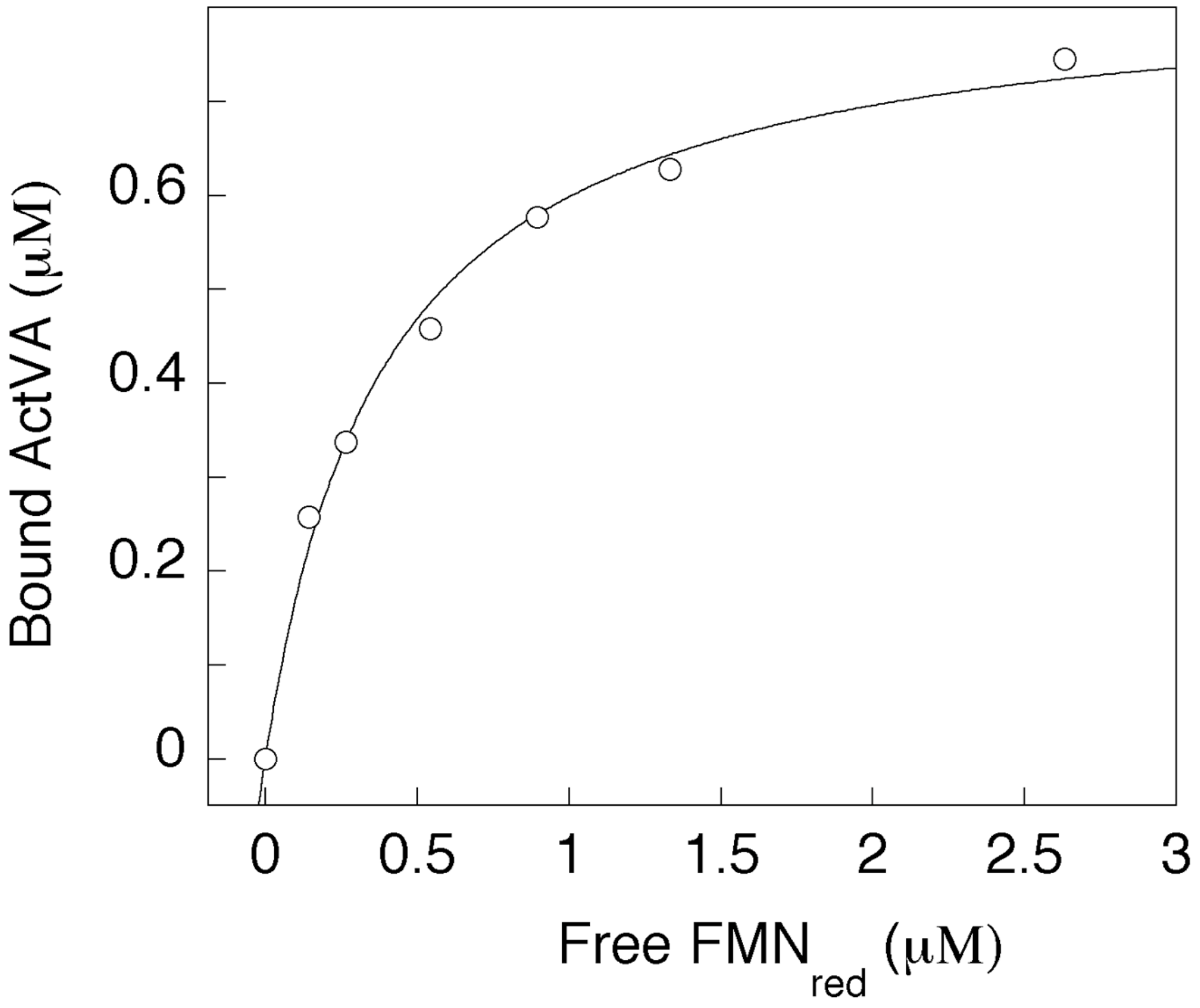

Determination of the $K_d$ value for $FMN_{red}$ with regard to ActVA. ActVA (1 μM) was incubated in anaerobiosis with increasing amounts of $FMN_{red}$ in 50 mM Tris-HCl pH7.6 at 18 °C. Sample was excited at 295 nm and the fluorescence spectra were recorded and the fluorescence intensity at 339 nm was used to determine the amount of $FMN_{red}$ bound to ActVA.

Supplementary data Figure 4

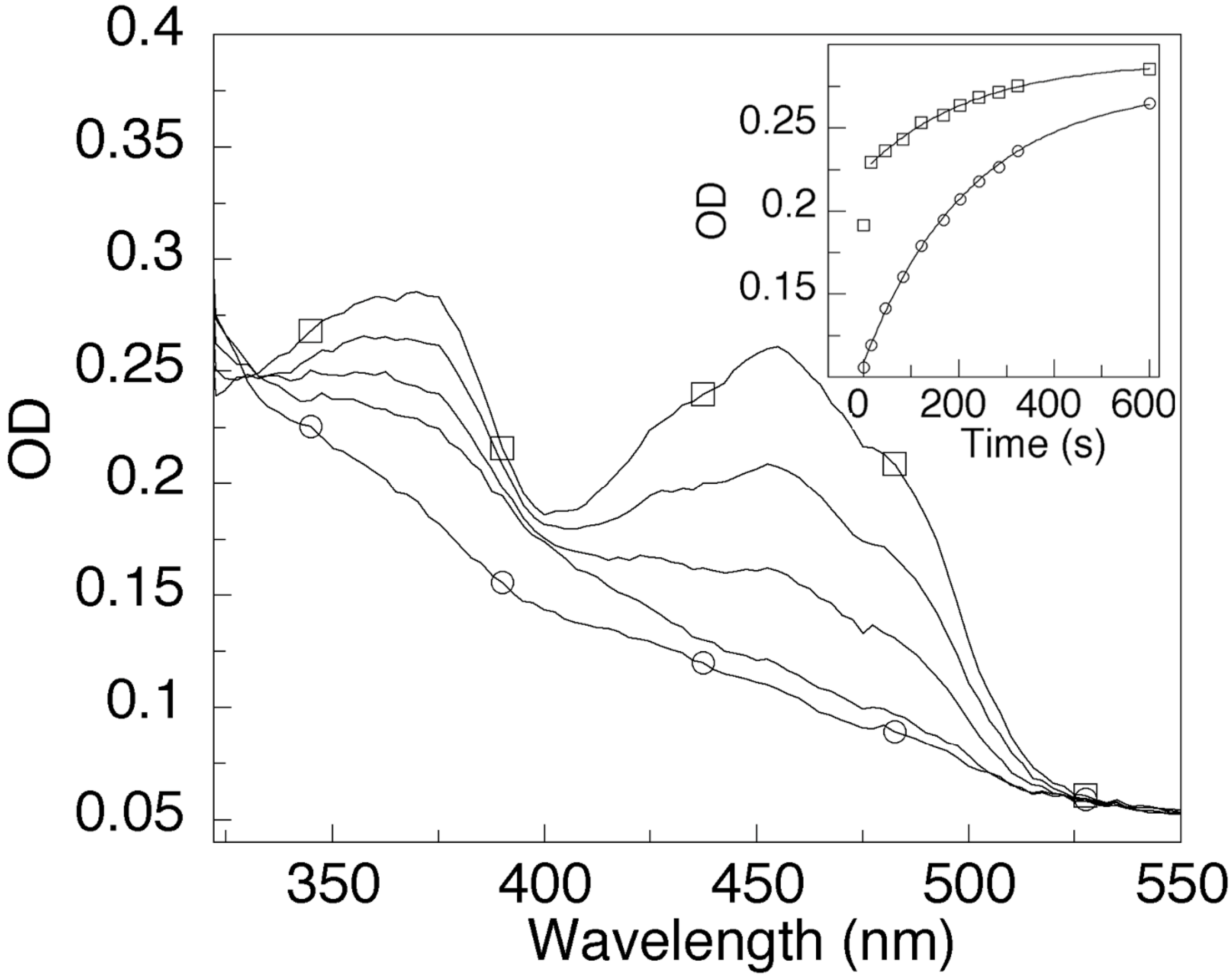

Oxidation of ActVB-$FMN_{red}$ in the presence of 1 equivalent of ActVA. ActVB (60 µM) containing $FMN_{ox}$ (16 µM) was anaerobically mixed with NADH (50 µM) in 10 mM Tris-HCl pH 7.6. After total reduction of $FMN_{ox}$ by NADH, ActVA (60 µM) was added. Then, 50 µM $O_2$ was rapidly added and UV-Visible spectra were recorded from the bottom to the top at 0 s (○), 15 s, 80 s, 200 s and 600 s (□). Inset shows the variation of absorbance at 380 nm (□) and 457 nm (○) as a function of time. Lines were calculated for best fit to a exponential model.